# Some Additional Bounds on the Photon Charge

C Sivaram and Kenath Arun

Indian Institute of Astrophysics, Bangalore, India

**Abstract:** We have arrived at tight constraints on the photon charge, giving comparable bounds, one based on the dominance by dark energy at the present epoch, and the other based on the requirement that early universe nucleosynthesis not be affected by any residual electrostatic energy due to any miniscule charge on the radiation photons in that era. Limits have also been arrived at from synchrotron and IC effects. We have also set limits on the charge based on the properties of black holes. The set of constraints arrived at in this paper are consistent with those predicted by other authors.

# 1. Introduction

In a recent paper[1] a stringent bound on the photon charge was put based on the phase coherence of extragalactic radiation, i.e., based on the fact that electromagnetic waves moving along different paths do not acquire phase difference. The bound put was < $10^{-32}$e. The particle data group[2] in 2006 listed only four bounds on the photon charge, although several limits have been put on the photon mass.

There have been earlier tight constraints on the photon charge based on very different considerations like the CMBR, gamma ray bursts[3, 4] and magnetic field deflections.[5, 6] Here we put tight constraints, giving comparable bounds, one based on the dominance by dark energy at the present epoch, and the other based on the requirement that early universe nucleosynthesis not be affected by any residual electrostatic energy due to any miniscule charge on the radiation photons in that era. The latter limit also implies a photon charge of less than $10^{-32}$e.

The above limits are independent of the photon mass. However if limits on photon mass[7, 8, 9] can also be used, then further constraints on the photon charge can be put from the possibility that in compact objects like supernova remnants (a neutron star) or gamma ray bursts, there could be profuse synchrotron and inverse Compton radiation from such particles in the presence of strong magnetic fields and radiation densities.

The current interest in millicharged particles[10, 11, 12] is also relevant in this context as limits can be put on their masses. Also neutrino charges[13, 14] can also be constrained from such considerations.

## 2. Constraint on photon charge from cosmological considerations

We can arrive at a constraint on the photon charge independently of its mass from cosmological considerations. The total number of photons in the CMB is of the order of $10^{90}$ and this number is conserved during the expansion of the universe[15]. If each of the photons has a miniscule charge $q$, then the total electrostatic energy due to the charge is given by:

$$\frac{N^2 q^2}{R_H} \qquad \ldots (1)$$

Where, $R_H \sim 10^{28}$ cm is the Hubble radius. The energy due to these charges will cause anisotropy in the universe. But its effect should be smaller than the observed anisotropy of 1 part in $10^5$ in the CMB. Hence:

$$\frac{N^2 q^2}{R_H} \approx 10^{-5} E_\gamma \qquad \ldots (2)$$

Where, $E_\gamma = aT_\gamma^4 (2\pi^2 R_H^3)$, and $2\pi^2 R_H^3$ is the Hubble volume and $T_\gamma \approx 2.7 K$. This sets a limit on the charge of the photon as: $q < 10^{-32} e$. [For example see ref. 3]

At the big bang nucleosynthesis (BBN) the temperature was of the order of a billion Kelvin, which is given by:[15]

$$T = \frac{10^{10}}{t^{1/2}} \qquad \ldots (3)$$

The corresponding energy density is given by: $\varepsilon = aT^4 \approx 10^{22}\, ergs/cc$

The expansion rate of the universe is given by:

$$\frac{\dot{R}}{R} \propto \sqrt{\varepsilon} \qquad \ldots (4)$$

If the energy density due to the photon charge, that is the electrostatic energy, is more than 10% of this, the expansion rate would then change the primordial element abundance.

The electrostatic energy of all the $\sim 10^{90}$ photons (as photon number is conserved) is:

$$\sim \frac{N^2 q^2}{R^4} < 10^{21}\, ergs/cc \qquad \ldots (5)$$

Where, $R$ is the universe scale factor corresponding to a temperature of a billion Kelvin (i.e., t~ few seconds), the nucleosynthesis era. This implies $q < 10^{-44}\, esu$, thus putting the limit on the photon charge as $q \sim 10^{-34} e$.

This is a very stringent limit comparable to the one in [ref 1] and also with [ref 6].

Of course the light neutrino number $\sim 10^{88}$ is also conserved, but also includes antineutrinos. So if the antineutrinos have opposite charges the similar limit on the neutrino electric charge (as that of the argument above) may not apply.

However in the following calculation for Synchrotron and Inverse Compton, the magnitude of the power (energy) emitted goes as the square of the charge and so the derived constraints apply, for both photons and neutrinos.

## 3. Constraint on photon charge from synchrotron and IC radiation

A charged particle moving in a magnetic field radiates energy. At non-relativistic velocities, this results in cyclotron radiation while at relativistic velocities it results in synchrotron radiation. The relativistic form of the equation of motion of a particle in a magnetic field is,

$$\frac{d}{dt}(\gamma m v) = q(v \times B) \qquad \ldots (6)$$

Where $v$ is the velocity vector of a particle of charge q, the magnetic and electric vectors are B and E, m is the mass, and gamma is the usual Lorentz factor.

The radiation emission for a relativistic electron moving in a magnetic field B is given by

$$-\dot{E} = \frac{dE}{dt} = \frac{2}{3}\frac{e^4}{m_0^2 c^3} B^2 \gamma^2 \qquad \ldots (7)$$

Where, $\gamma = \frac{E}{m_e c^2}$

The total power for an isotropic distribution of synchrotron radiation is given by:

$$P = \frac{4}{3}\sigma_T c \beta^2 \left(\frac{B^2}{8\pi}\right)\gamma^2 \qquad \ldots (8)$$

Inverse Compton scattering occurs when a relativistic electron scatters a low energy photon to a higher energy. In the case of IC effect, the power radiated by each electron is given by,

$$-\dot{E} = \frac{dE}{dt} = \frac{2}{3}\frac{e^4}{m_0^2 c^3} \varepsilon \gamma^2 \qquad \ldots (9)$$

Where $\varepsilon = aT^4$ is the photon flux density and $a = 7.5657 \times 10^{-15} \, ergs/cm^3/K^4$.

Limits for the charges can be obtained from the synchrotron and IC emission from compact objects having high magnetic fields (>$10^{12}$ T) and high temperature at time of formation (~$10^{12}$ K), as these objects are involved in high energy transient events like gamma ray burst and supernova.

Tight constraints arise because of the ($1/m^4$) dependence in both cases, which can be seen from (7) and (9). Photons and neutrino rest masses are expected to be very tiny, especially photon rest mass <$10^{-15}$ eV.

We also obtain tight constraints on masses of millicharge particles[10, 11, 12]. The total amount of energy of synchrotron and IC cannot exceed the energy in the outburst (the upper limit of ~$10^{51}$ ergs).

From the synchrotron and IC radiation loss, we can set a limit on the mass of neutrinos and photons since the power due to these effects is constrained by the fact that this energy should be less than the total energy liberated during the process, which is of the order of $10^{51}$ ergs.

Neutrinos:

| Synchrotron | IC |
|---|---|
| B~$10^{12}$ T | T~$10^{12}$ K |
| m ~ 1 GeV | m ~ 10 GeV |

Photons:

| Synchrotron | IC |
|---|---|
| B~$10^{12}$ T | T~$10^{12}$ K |
| m ~ 0.1 GeV | m ~ 10 GeV |

Since the masses are too large they should have already been seen in cosmic rays and accelerators, which is not the case. Hence we can rule out the possibility of millicharges.

The mass of the photon is thought to be of the order of $10^{-15}$ eV and charge of $10^{-30}e$.[1] From equations (7) and (9), the power due to synchrotron and IC, for these values, is of the order of $10^{31}$ ergs/s and $10^{41}$ ergs/s respectively. The upper limit on the power radiated during the process sets a limit on the charge of the photon as $10^{-28}e$, which matches with the limits predicted by other authors through other indirect deductions.

Similarly, for the neutrinos of mass $10^{-5}$ eV and charge of $10^{-30}e$, the power radiated during synchrotron and IC are of the order of $10^{-9}$ ergs/s and 10 ergs/s. the limit on the neutrino charge based on the upper limit of the power radiated during these processes is of the order of $10^{-18}e$.

**4. Constraint on photon charge from black hole properties**

Further, it is interesting that we can put similar limits from the well known properties of charged black holes[16]. In particular, charged black holes have an upper limit on the electric charge they can have which is related to their masses. The event horizon for these charged black holes is given by:

$$r = m \pm (m^2 - q^2)^{1/2} \qquad \ldots (10)$$

This implies that $q \leq m$ in the natural units, where $G = c = h = 1$, or $q \leq \sqrt{G}m$

For a solar mass black hole, this limit works out to be of the order of:

$$q \leq \sqrt{6.67 \times 10^{-8}} \times (2 \times 10^{33}) = 5 \times 10^{29}\, esu \Rightarrow q \leq 10^{39} e \qquad \ldots (11)$$

If the solar mass black hole is formed by the collapse of 1 eV photons, then the number of particles in the black hole is given by:

$$N \approx \frac{10^{33} g}{10^{-33} g} \approx 10^{66} \qquad \ldots (12)$$

Hence the charge associated with each of these photons works out to be of the order of

$$q = \frac{10^{39} e}{10^{66}} = 10^{-27} e \qquad \ldots (13)$$

The above obtained limit on the photon charge matches with the limits obtained from other independent considerations and also matches with the limits obtained by other authors.

Similarly in the case of supermassive black holes ($10^9$ solar mass), the limit works out to be of the order of $10^{-28}$e. This implies that for the formation of the black hole, it should be electrically neutral to one part in $10^{18}$.

### 5. Conclusion

In this paper we have arrived at a set of constraints on the electric charge of the photon based on the cosmological considerations involving dark energy and the fact that any miniscule charges on the radiation photons should not affect the early universe BBN. Both these constraints are independent of the mass of the photon. By assuming the constraints on the mass of the photons, suggested by various authors, we have set bounds on their charge from radiations emitted in the presence of strong magnetic fields. We have also set limits on the charge based on the properties of black holes. The set of constraints arrived at in this paper are consistent with each other and also matches with those predicted by other authors.